\documentclass[prd,twocolumn,showpacs,amsmath,amssymb]{revtex4}
\usepackage{graphicx,color,slashbox}
\begin{document}
%\begin{CJK}{GBK}{song}

\title{A Possibility of Search for New Physics at LHCb}

\vspace{1cm}

\author{Xiang Liu$^1$}

\author{Hong-Wei Ke$^2$}
\author{Qing-Peng Qiao$^2$}
\author{Zheng-Tao Wei$^2$}
\author{Xue-Qian Li$^2$}
\vspace{1cm}

\affiliation{$^1$ Department of Physics, Peking University,
100871, Bejing, China,
\\ $^2$ Department of Physics, Nankai University, 300071, Tianjin, China}

\vspace{2cm}

%\date{\today}

\begin{abstract}

It is interesting to search for new physics beyond the standard
model at LHCb. We suggest that weak decays of doubly charmed baryon
such as $\Xi_{cc}(3520)^+,\;\Xi_{cc}^{++}$ to charmless final states
would be a possible signal for new physics. In this work, we
consider two models, i.e. the unparticle and $Z'$ as examples to
study such possibilities. We also discuss the cases for
$\Xi^0_{bb},\;\Xi_{bb}^-$ which have not been observed yet, but one
can expect to find them when LHCb begins running. Our numerical
results show that these two models cannot result in sufficiently
large decay widths, therefore if such modes are observed at LHCb,
there must be a new physics other than the unparticle or $Z'$
models.

\end{abstract}


\maketitle

\section{Introduction}\label{sec1}

LHC will begin its first run pretty soon, and besides searching for
the long-expected Higgs boson, its main goal is to explore new
physics beyond the SM. Many schemes have been proposed to reach the
goal. Indeed, the LHCb detector, even though is not responsible for
the Higgs hunting, will provide an ideal place to study heavy flavor
physics and search for evidence of new physics. One can make careful
measurements on rare decays of B-mesons, b-baryons, B-mixing and CP
violation with a huge database available at LHCb, moreover, we are
inspired by the possibilities of discovering new physics. It would
be beneficial to conjecture more possible processes which would
signal existence of new physics.

%In this work, we propose that measuring the decay rates of
%double-charmed baryons and double-bottomed baryons may be very
%helpful for exploring new physics.

In 2002, the first event for doubly charmed baryon,
$\Xi^{+}_{cc}(3520)$, was observed by the SELEX Collaboration in the
channel of $\Xi_{cc}^{+}\to \Lambda_{c}^{+}K^{-}\pi^{+}$
\cite{SELEX-DOUBLE CHARM,SELEX-DOUBLE-1,SELEX-DOUBLE-2}.
$\Xi^{+}_{cc}$ has the mass $m=3519\pm1$ MeV and width $\Gamma<5$
MeV. By studying an alternative channel of $pD^{+}K^{-}$ conducted
later, the mass of the baryon-resonance was confirmed as $m=3518\pm
3$ MeV \cite{SELEX-confirm-double}, which is consistent with that
given in Ref. \cite{SELEX-DOUBLE CHARM}. In the present theory,
there definitely is no reason to exclude existence of
$\Xi_{cc}^{++}$ which contains $ccu$ valence quarks and as well
$\Xi_{bb}^0 \;(bbu)$ and $\Xi_{bb}^- \; (bbd)$, by the flavor-SU(3)
symmetry.

%The fixed target experiments cannot provide a sufficiently large
%database for more accurate measurements, especially for the rare
%decays. By contrast, a remarkable amount of such baryons will be
%produced at LHCb, and the available data would enable physicists
%to search for new physics signals.
In this work, we propose that direct decays of $\Xi_{cc}$ with
charmless final states or $\Xi_{bb}$ with bottomless final states
would be signals for new physics. By the quark-diagrams, one can
easily notice that the main decay modes of $\Xi_{cc}$ would be
$D^+\Lambda(\Sigma^0)$, $\Lambda_c K^0$, $D^+ PK^-$ and
$\Lambda_cK^-\pi^+$. The later two modes are just the channels where
the SELEX collaboration observed the baryon $\Xi_{cc}$.
%Then the charmed hadrons decay into non-charmed final states which are
%eventually observed at detectors. There are two successive processes
%and both of them are realized via weak interactions. Generally, such
%decay modes possess final states with multi-hadrons and the
%mechanisms responsible for those reactions are the standard model.
While the direct decays of $\Xi_{cc}$ (or $\Xi_{bb}$) into charmless
(bottomless) final states are suppressed in the standard model, so
that would be sensitive to new physics beyond the SM.

Since in $\Xi_{cc}$ there are two identical charm quarks which can
neither annihilate, nor exchange W-boson to convert into other
quarks. In the SM, direct transition of $\Xi_{cc}$ into charmless
final states may realize via the double-penguin mechanism which is
shown in Fig. 1 (a), the crossed box-diagram (Fig. 1 (b)) and a
possible two-step process shown in Fig. 1 (c). The mechanism
includes two penguin loops or a crossed box-diagram is very
suppressed, so that cannot result in any observable effects and we
can ignore them completely.  If a non-zero rate is observed at LHCb,
it should be a signal of new physics. Definitely the  diagram of
Fig. 1 (c) may cause a non-zero contribution and contaminate our
situation for exploring new physics. If we consider the charmless
decays of $\Xi_{cc}^{++}$ or bottomless decays of $\Xi_{bb}^-$, that
diagram (Fig. 1 (c)) does not exist at all. Then, the first question
is that can we distinguish such direct decays of $\Xi_{cc}$ into
charmless final states (or $\Xi_{bb}$ into bottomless final states)
from the secondary decays which result in charmless (or bottomless)
products and are the regular modes in the framework of the SM. The
answer is that the direct transitions are favorably two-body decays,
namely in the final states there are only two non-charmed hadrons by
whose momenta one can re-construct the invariant mass spectra of
$\Xi_{cc}$ (or $\Xi_{bb}$), whereas, in the regular modes with
sequent decays, there are at least three hadrons in the final
states.
\begin{figure}[t]
\begin{center}
\begin{tabular}{ccc}
\scalebox{0.6}{\includegraphics{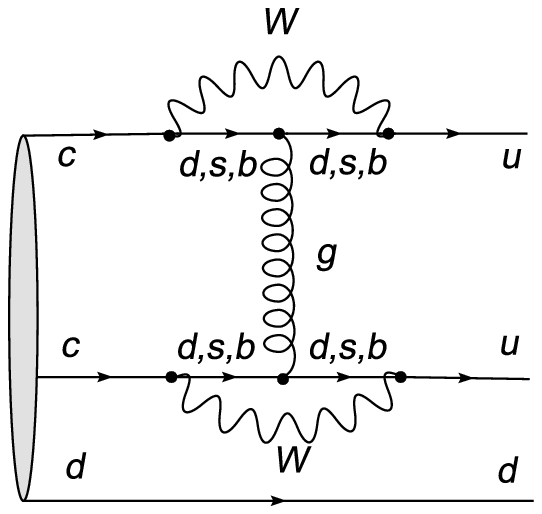}}\\(a)\\\scalebox{0.7}{\includegraphics{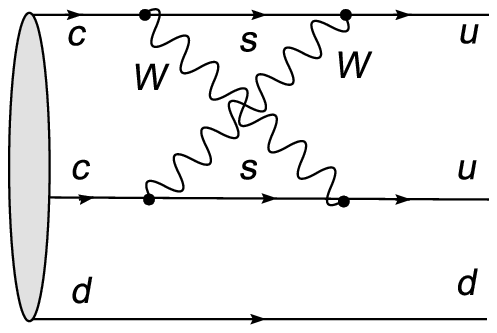}}\\
(b)\\\scalebox{0.4}{\includegraphics{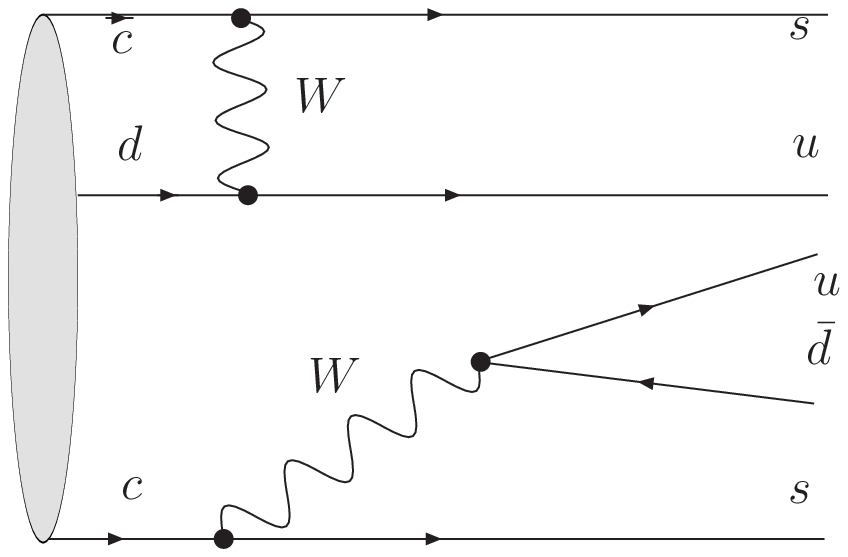}}\\(c)
\end{tabular}
\end{center}
\caption{(a) The double-penguin diagram which can induce the decay
of $\Xi_{cc}(\Xi_{bb})$ into non-charm (non-bottom) final states.
(b) The crossed box diagram. (c) An emission where the effective
interaction would be non-local and for charmless decays of
$\Xi_{cc}^{++}$, it does not exist. \label{SM}}
\end{figure}

The second question is that is there any mechanism beyond the
standard model available which can result in such direct decays?
Below, we use two models to demonstrate how such direct decay modes
are induced and estimate the widths accordingly. One of them is the
unparticle scenario and another one is the $SU(3)\times
SU(2)_L\times SU(2)_R \times U(1)_{B-L} $ model where a new gauge
boson $Z'$ exists and mediates an interaction to turn the charm
quark into a $u$-quark. Thus by exchange of an unparticle or $Z'$
between the two charm quarks in $\Xi_{cc}^{+(++)}$ (or between the
two bottom quarks in $\Xi_{bb}^{-(0)}$), these direct transitions
occur.

In this work, for simplicity, we only consider the inclusive decays
of $\Xi_{cc}^{++}\; (\Xi_{bb}^-)$ into charmless (bottomless) final
states. The advantage of only considering the inclusive processes is
obvious that we do not need to worry about the hadronization of
quarks into final states because such processes are fully governed
by the non-perturbative QCD effects and brings up much uncertainty.

Below, we will investigate the processes caused by exchanging
unparticles and $Z'$ separately and then make a brief discussion on
the possibility that new physics may result in observable phenomena
at LHCb.

\section{The inclusive decay of doubly charmed baryon }\label{sec2}

\subsection{The unparticle scenario}

Before entering the concrete calculation, we briefly review the
concerned knowledge on the unparticle physics \cite{theory}, which
is needed in later derivation. The effective Lagrangian describing
the interaction of the unparticle with the SM quarks is
\begin{eqnarray}
\mathcal{L}&=&\frac{c_S^{qq'}}{\Lambda_{\mathcal{U}}^{d_{\mathcal{U}}}}\bar{q}\gamma_{\mu}
(1-\gamma_5)q'\partial^{\mu}O_{\mathcal{U}}\nonumber\\
&&+\frac{c_V^{qq'}}{\Lambda_{\mathcal{U}}^{d_{\mathcal{U}}-1}}
\bar{q}\gamma_{\mu} (1-\gamma_5)q'O_{\mathcal{U}}^{\mu}+h.c.,
\end{eqnarray}
where $O_{\mathcal{U}}$ and $O_{\mathcal{U}}^{\mu}$ are the scalar
and vector unparticle fields respectively. $q$ and $q'$ denote the
SM quark fields. Generally, the dimensionless coefficients
$c_{_{S,V}}^{qq'}$ is related to the flavor of the quark field.
%but in this work, we set $c_{_S}^{qq'}=c_{_S}$ and
%$c_{_V}^{qq'}=c_{_V}$ for simplifying the phenomenological
%analysis.
This interaction induces a FCNC and contributes to the processes of
concern.

For a scalar unparticle field, the propagator with momentum $p$ and
scale dimension $d_{\mathcal{U}}$ is \cite{propagator}
\begin{eqnarray}
&&\int d^4 x d^{ip\cdot
x}\langle0|TO_{\mathcal{U}}(x)O_{\mathcal{U}}(0)|0\rangle\nonumber\\
&&=i\frac{A_{d_{\mathcal{U}}}}{2\sin(d_{\mathcal{U}}\pi)}
\frac{1}{(-p^2-i\epsilon)^{2-d_{\mathcal{U}}}}
\end{eqnarray}
with
\begin{eqnarray}
A_{d_{\mathcal{U}}}=\frac{16\pi^{5/2}}{(2\pi)^{2d_{\mathcal{U}}}}
\frac{\Gamma(d_{\mathcal{U}}+1/2)}{\Gamma(d_{\mathcal{U}}-1)\Gamma(2d_{\mathcal{U}})},
\end{eqnarray}
with $d_{\mathcal{U}}$ the scale dimension.

For the vector unparticle, the propagator reads
\begin{eqnarray}
&&\int d^4 x e^{ip\cdot x}\langle0|T
O_{\mathcal{U}}^{\mu}O_{\mathcal{U}}^{\nu}(0)|0\rangle\nonumber\\
&&=i\frac{A_{d_{\mathcal{U}}}}{2\sin(d_{\mathcal{U}}\pi)}
\frac{-g^{\mu\nu}+p^{\mu}p^{\nu}/p^2}{(-p^2-i\epsilon)^{2-d_{\mathcal{U}}}},
\end{eqnarray}
where the transverse condition
$\partial_{\mu}O_{\mathcal{U}}^{\mu}=0$ is required.

\begin{figure}[htb]\caption{The inclusive transition of doubly charmed baryon in
unparticle physics, where the double-dashed line denotes the
scalar or vector unparticle in the unparticle model or $Z'$ in the
left-right model. \label{decay}}
\begin{center}
\scalebox{1}{\includegraphics{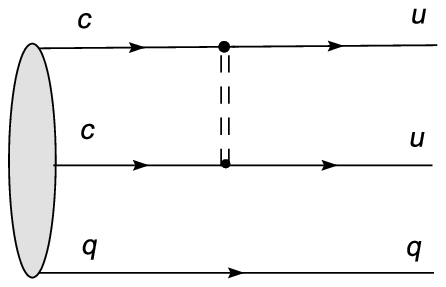}}\\(a)\\
\scalebox{0.5}{\includegraphics{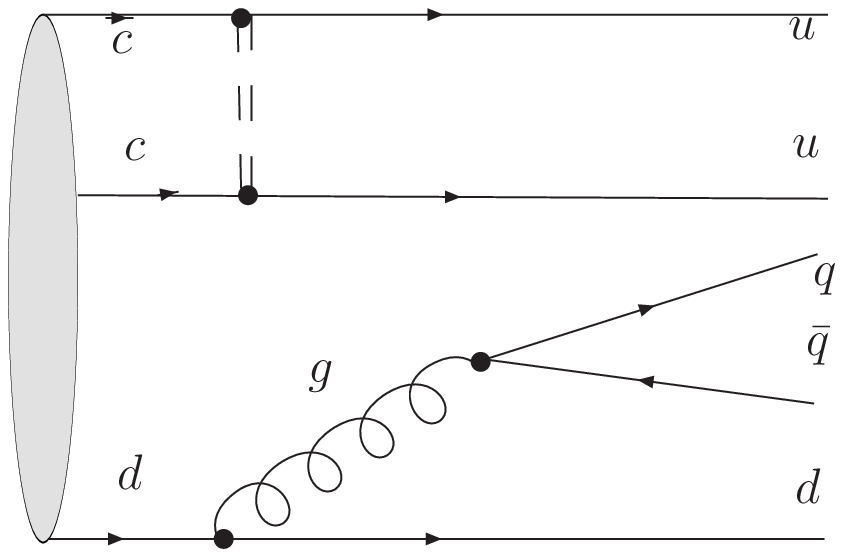}}\\(b)
\end{center}

\end{figure}

In the unparticle physics, the inclusive decay of doubly charmed
baryons into light quarks $ccq\to uuq$ occurs at the tree level,
and the transition is depicted in Fig. \ref{decay}. Here the
exchanged agent between the two charm quarks can be either scalar
or vector unparticle.

Even though we only consider the inclusive processes where the
quarks in the final states are treated as on-shell free particles
and the wavefunctions of the light hadrons in the final states are
not needed, the binding effect of the initial baryons
($\Xi_{cc}^{++}$ or $\Xi_{bb}^-$) which are composed of three
valence quarks must be taken into account. Namely, when we
calculate the hadronic matrix elements, we need to invoke concrete
phenomenological models to carry out the computations where the
wave function of the initial baryon is needed. In this work, we
adopt a simple non-relativistic model, i.e. the harmonic
oscillator model \cite{harmolic oscillator}. This model has been
widely and successfully employed in similar researches
\cite{HO-prove-1,HO-prove-2,HO-prove-3,HO-prove-4,HO-prove-5,HO-prove-6,HO-prove-7}.
Thus one can trust that for heavy hadrons, such simple
non-relativistic model can work well and the results are
relatively reliable, even though certain errors are not avoidable.
Thus in this work the matrix elements of the effective operators
evaluated in terms of the harmonic oscillator wave function is
believed to be a good approximation. According to the references
listed above, the errors in the estimate, especially as we only
need the wavefunction of the the initial hadron, are expected to
be less than 10\%. By changing the input parameters and the model
parameters which are obtained by fitting other experiments, we
scan the region of changes of the numerical results and find that
the error range is indeed consistent with our expectation.

In the harmonic oscillator model, the wave function of the doubly
charmed baryon $\Xi_{cc}(3520)^+$ is expressed as
\begin{eqnarray*}
&&|\Psi_{\Xi_{cc}^+}(P,s)\rangle\nonumber\\&=&\mathcal{N}
\sum_{color,spin}\chi_{\mathcal{_{SF}}}\varphi_{\mathcal{_C}}
\int d^{3}p_{\rho}d^{3}p_{\lambda}\nonumber\\&&
\times\psi_{\Xi_{cc}^+}(\mathbf{p}_{\rho},\mathbf{p}_{\lambda})
b^{\dag}_{c}(p_1',s_1')b^{\dag}_{c}(p_2',s_2')b^{\dag}_{u}(p_3',s_3')|0\rangle,
\end{eqnarray*}
which satisfies the normalization condition
\begin{eqnarray*}
&&\langle\Psi_{\Xi_{cc}^+}(\mathbf{\mathbf{P}},s)|\Psi_{\Xi_{cc}^+}(\mathbf{P}',s')\rangle
\nonumber\\&=&(2\pi)^3\frac{M_{\Xi_{cc}^+}}{E_P}\delta^3(\mathbf{P}-\mathbf{P}')\delta(s-s'),
\end{eqnarray*}
where $\mathcal{N}$ is the normalization constant.
$\chi_{\mathcal{_{SF}}}$ and $\varphi_{\mathcal{_C}}$ denote the
spin-flavor and color parts of the wavefunction of doubly charmed
baryon $\Xi_{cc}(3520)^+$ respectively whose explicit expressions
are
 \begin{eqnarray*}
\mathcal{N}&=&\sqrt{\frac{E }{M}\frac{m_1' m_2' m_3'}{ E_1' E_2'
E_3'}},\;\;\;\varphi_{\mathcal{_C}}=\frac{1}{\sqrt{6}}\epsilon_{ijk},\\
\chi_{\mathcal{_{SF}}}&=&\frac{1}{\sqrt{6}}\Big[2|c\uparrow
c\uparrow d\downarrow\rangle-|c\uparrow c\downarrow
d\uparrow\rangle-|c\downarrow c\uparrow d\uparrow\rangle\Big].
\end{eqnarray*}
In the harmonic oscillator model, the spatial wavefunction
$\psi_{\Xi_{cc}^+}(\mathbf{p}_{\rho},\mathbf{p}_{\lambda})$ reads
as
\begin{eqnarray*}
&&\psi_{\Xi_{cc}^+}(\mathbf{p}_{\rho},\mathbf{p}_{\lambda})\\&=&3^{3/4}\Big(\frac{1}{\pi
\alpha_{\rho}^2}\Big)^{3/4}\Big(\frac{1}{\pi
\alpha_{\lambda}^2}\Big)^{3/4}\exp\Big[-\frac{\mathbf{p}_{\rho}^2}
{2\alpha_{\rho}}-\frac{\mathbf{p}_{\lambda}^2}{2\alpha_{\lambda}}\Big]
\end{eqnarray*}
with the definitions
\begin{eqnarray*}
\mathbf{p}_\rho&=&\frac{\mathbf{p}_1'-\mathbf{p}_2'}{\sqrt{2}},\;\;\;
\mathbf{p}_\lambda=\frac{\mathbf{p}_1'+\mathbf{p}_2'-\frac{2m_c}{m_d}
\mathbf{p}_3'}{\sqrt{2\frac{2m_c+m_d}{m_d}}},\\
\mathbf{P}&=&\mathbf{p}_1'+\mathbf{p}_2'+\mathbf{p}_3',
\end{eqnarray*}
and the parameters $\alpha_\rho$ and $\alpha_\lambda$ reflect the
non-perturbative effects and will be given in later subsection.

In the center of mass frame of $\Xi_{cc}(3520)^+$, the hadronic
matrix elements $S_{fi}$ is written as
\begin{eqnarray*}
S_{fi}&=&(2\pi)^4\delta^4(p_1+p_2+p_3-M)T
\end{eqnarray*}
with $T= (T_S+T_V)$.
%with
%\begin{eqnarray*}
%T&=& \int d^{3}\mathbf{p}_{\rho}d^{3}\mathbf{p}_{\lambda}
%(2\pi)^3\frac{E_{p_3}}{m_{p_3}}[\bar{u}^{i}_{u}(p_1,s_1)\Gamma_1{u}^{ia}_{c}(p_1',s_1')
%\nonumber\\&&\times\bar{u}^{j}_{u}(p_2,s_2)\Gamma_2
%{u}^{j}_{c}(p_2',s_2') iD(q)]
%[\mathcal{N}\sum_{spin}\psi_{ccd}(\mathbf{p}_{\rho},\mathbf{p}_{\lambda})].
%\end{eqnarray*}

For exchanging scalar unparticle, $T_S$ matrix element is written
as
\begin{eqnarray}
T_{S}&=&\sum_{spin} \int d^{3}{p}_{\rho}d^{3}{p}_{\lambda}
(2\pi)^3\frac{E_{p_3}}{m_{p_3}}\nonumber\\&&\times[\bar{u}_{u}(p_1,s_1)
\gamma_\mu(1-\gamma_5){u}_{c}(p_1',s_1')
\nonumber\\&&\times\bar{u}_{u}(p_2,s_2)\gamma_\nu(1-\gamma_5)
{u}_{c}(p_2',s_2') ]\nonumber\\&&\times
(\frac{c_S^{cu}}{\Lambda_\mathcal{U}^{d_\mathcal{U}}})^2\frac{A_{d_{\mathcal{U}}}}
{2\sin(d_{\mathcal{U}}\pi)} \frac{iq^\mu
q^\nu}{(-p^2-i\epsilon)^{2-d_{\mathcal{U}}}}\nonumber\\&&\times
\mathcal{N}\psi_{\Xi_{cc}^+}(\mathbf{p}_{\rho},\mathbf{p}_{\lambda}).
\end{eqnarray}
For the vector unparticle exchange, $T_V$ is
\begin{eqnarray}
T_V&=&\sum_{spin}\int d^{3} {p}_{\rho}d^{3} {p}_{\lambda}
(2\pi)^3\frac{E_{p_3}}{m_{p_3}}\nonumber\\&&\times[\bar{u}_{u}(p_1,s_1)
\gamma_\mu(1-\gamma_5){u}_{c}(p_1',s_1')
\nonumber\\&&\times\bar{u}_{u}(p_2,s_2)\gamma_\nu(1-\gamma_5)
{u}_{c}(p_2',s_2') ]\nonumber\\&&\times
(\frac{c_V^{cu}}{\Lambda_\mathcal{U}^{d_\mathcal{U}-1}})^2\frac{A_{d_{\mathcal{U}}}}
{2\sin(d_{\mathcal{U}}\pi)} \frac{i(-g^{\mu\nu}+q^\mu
q^\nu/q^2)}{(-p^2-i\epsilon)^{2-d_{\mathcal{U}}}}
\nonumber\\&&\times\mathcal{N}\psi_{\Xi_{cc}^+}(\mathbf{p}_{\rho},\mathbf{p}_{\lambda}).
\end{eqnarray}
Here $u_q$ and $\bar{u}_q$ ($q=c,u$) denote the Dirac spinors
\begin{eqnarray}
u_q&=&\sqrt{\frac{E_q+m_q}{2m_q}}\left (
\begin{array}{c}
1
\\
\frac{\sigma\cdot p}{E_q+m_q}
\end{array} \right )\chi,
\\
\bar u_q&=&\sqrt{\frac{E_q+m_q}{2m_q}}\chi^\dagger\large\left(1
,\,\,\,\,-\frac{\sigma\cdot p}{E_q+m_q}\large\right),
\end{eqnarray}
and we can use the expression \cite{Li}
\begin{eqnarray}
\frac{|c_{s}^{cu}|^2}{\Lambda^{2d_\mathcal{U}}_\mathcal{U}}=\frac{6m\,
\Delta m |\sin d_{ \mathcal{U}} \pi|}{5 f^2
\hat{B}A_{d_\mathcal{U}}m^{2d_\mathcal{U}}},\\
\frac{|c_{s}^{cu}|^2}{\Lambda^{2d_\mathcal{U}-1}_\mathcal{U}}=\frac{2m\,
\Delta m |\sin d_{ \mathcal{U}} \pi|}{ f^2
\hat{B}A_{d_\mathcal{U}}m^{2d_\mathcal{U}-2}},
\end{eqnarray}
to simplify $T_V$ and $T_S$. One needs to sum over all possible
spin assignments for the Dirac spinors.

\subsection{The $Z'$ scenario}

The Left-Right models \cite{LR} is also a natural extension of the
electroweak model. It has been widely applied to the analysis on
high energy processes. For example, recently He and Valencia
\cite{h1} employed this model with certain modifications to
explain the anomaly in  $A_{FB}^b$ observed at LEP \cite{exp1}.
Barger et al. studied $Z'$ mediated flavor changing neutral
currents in $B$-meson decays \cite{B-decays-z}, $B_s-\bar{B}_s$
mixing \cite{Bmixing-z} and $B\to K\pi$ puzzle \cite{bkpi}.

The gauge group of the model \cite{h1} is $SU(3)\times SU(2)_L\times
SU(2)_R \times U(1)_{B-L} $ where the four gauge couplings $g_3 ,\;
g_L ,\; g_R $ and $g$ correspond to the four sub-groups
respectively. The vacuum expectation values of the three Higgs
bosons break the symmetry. The symmetry breaking patterns are
depicted in literature. The introduction of a scalar field $\phi$
causes $Z_0$ in the standard model to mix with a new gauge boson
$Z_R$, then $Z,\; Z'$ are the mass eigen-states.

For the neutral sector the Larangian is
\begin{eqnarray}
\mathcal{L}&=&-\frac{g_L}{2\rm{cos}\theta_W}\bar{q}\gamma^{\mu}(g_V-g_A\gamma_5)q
(\rm{cos}\xi_ZZ_\mu-\rm{sin}\xi_ZZ'_\mu)\nonumber\\&&
+\frac{g_Y}{2}{\rm{tan}}\theta_R(\frac{1}{3}\bar{q}_L\gamma^\mu
q_L+\frac{4}{3} \bar{u}_{Ri}\gamma^\mu
u_{Ri}-\frac{2}{3}\bar{d}_{Ri}\gamma^\mu
d_{Ri})\nonumber\\&&(\rm{sin}\xi_ZZ_\mu-\rm{cos}\xi_ZZ'_\mu)
\nonumber\\&&
+\frac{g_Y}{2}(\rm{tan}\theta_R+\rm{cot}\theta_R)(sin\xi_ZZ_\mu-cos\xi_ZZ'_\mu)\nonumber\\&&
(V^{d*}_{Rbi}V^{d}_{Rbj}\bar{d}_{Ri}\gamma^\mu
d_{Rj}-V^{u*}_{Rti}V^{u}_{Rtj}\bar{u}_{Ri}\gamma^\mu u_{Rj}).
\end{eqnarray}
Here $\theta_W$ is the electroweak mixing angle (${\rm{\tan
}\theta_W}=\frac{g_Y}{g_L}$), $\theta_R$ parameterizes the
relative strength of the right-handed interaction(${\rm{tan
}\theta_R}=\frac{g}{g_R}$), $\xi_Z$ is the $ Z - Z'$ mixing angle
and $V ^{u,d} _{Rij}$ are two unitary matrices that rotate the
right-handed up-(down)-type quarks from the weak eigen-states to
the mass eigen-states. Note that we use current notation for
Pati-Salam model, and only third family couples to $SU(2)_R$ in
this model.

In the $Z'$ model, inclusive decay of doubly charmed baryons into
light quarks $ccq\to uuq$ occurs  at tree level. The Feynman
diagram (Fig. \ref{decay}) is the same as that for the unparticle
scenario, but only the exchanged agent is replaced by $Z'$.

In the center of mass frame of  $\Xi_{cc}^+$, we can obtain
\begin{eqnarray}
T&=&\sum_{spin}\int d^{3} {p}_{\rho}d^{3} {p}_{\lambda}
(2\pi)^3\frac{E_{p_3}}{m_{p_3}}\nonumber\\&&\times[\bar{u}_{u}(p_1,s_1)
\gamma_\mu(1-\gamma_5){u}_{c}(p_1',s_1')
\nonumber\\&&\times\bar{u}_{u}(p_2,s_2)\gamma_\nu(1-\gamma_5)
{u}_{c}(p_2',s_2') ]\nonumber\\&&\times
\frac{-ig^{\mu\nu}}{(p^2+M_Z'^2)}
[\frac{g_L\rm{tan}\theta_W(\rm{tan}\theta_R+\rm{cot}\theta_R)cos\xi_Z}{2}
\nonumber
\\&&\times V^{u*}_{Rti}V^{u}_{Rtj}]^2\mathcal{N}\psi_{\Xi_{cc}^+}
(\mathbf{p}_{\rho},\mathbf{p}_{\lambda}).
\end{eqnarray}

The authors of Ref. \cite{h1,h2,h3} suggested that
$\rm{cot}\theta_R$ is large, so that $\rm{\tan}\theta_R$ can be
ignored. They took approximations
$\rm{tan}\theta_W\rm{cot}\theta_R\frac{M_W}{M_Z'}\sim 1$ and
$\rm{cos}\xi_Z\sim 1$. Because $M_{Z'}$ is larger than 500 GeV
\cite{h1}, one has $\frac{1}{p^2+M_Z'^2}\sim\frac{1}{M_Z'^2}$.

Then we have the final expression as
\begin{eqnarray}
T&=&\sum_{spin}\int d^{3} {p}_{\rho}d^{3} {p}_{\lambda}
(2\pi)^3\frac{E_{p_3}}{m_{p_3}}\nonumber\\&&\times[\bar{u}_{u}(p_1,s_1)
\gamma_\mu(1-\gamma_5){u}_{c}(p_1',s_1')
\nonumber\\&&\times\bar{u}_{u}(p_2,s_2)\gamma_\nu(1-\gamma_5)
{u}_{c}(p_2',s_2') ]\nonumber\\&&\times (-ig^{\mu\nu})G_F\sqrt{2}
(  V^{u*}_{Rti}V^{u}_{Rtj})^2\nonumber
\\&&\times\mathcal{N}\psi_{\Xi_{cc}^+}(\mathbf{p}_{\rho},\mathbf{p}_{\lambda}).
\end{eqnarray}

\subsection{The
expression of the decay width}

The inclusive decay rate would be obtained by integrating over the
phase space which involves three free quarks and the procedure is
standard \cite{3-body},
\begin{eqnarray}
&&\Gamma(\Xi_{cc}(3520)^+\rightarrow
u\,\,u\,\,d)\nonumber\\&&=\int^{a_{2}}_{a_{1}}
\mathrm{d}p_1^0\int^{b_{2}}_{b_{1}}
\mathrm{d}p_{2}^0\int^{2\pi}_{0} \mathrm{d}\eta
\int^{1}_{-1}\mathrm{d}(\cos\theta) \frac{
|T|^{2}}{16M_{{\Xi_{cc}^+} }(2\pi)^{4}},\nonumber\\
\end{eqnarray}
where $a_1$, $a_2$, $b_{1}$ and $b_2$ are defined as respectively
\begin{eqnarray*}
a_{1}&=&0,\;\;\; a_{2}=\frac{{M_{{\Xi_{cc}^+} }}}{2}- \frac{ {(
{m_{2}} + {m_{3}} ) }^2-m_1^2}
   {2\,{M_{{\Xi_{cc}^+} }}},\\
b_{1}&=&\frac{1}{2\,\tau}[\sigma(\tau+m_{+}m_{-})-\sqrt{{p_1^0}^2(\tau-m_{+}^2)(\tau-m_{-}^2)}],\\
b_{2}&=&\frac{1}{2\,\tau}[\sigma(\tau+m_{+}m_{-})+\sqrt{{p_1^0}^2(\tau-m_{+}^2)(\tau-m_{-}^2)}],\\
\sigma&=&M_{{\Xi_{cc}^+} }-p_1^0,\;\;\;
\tau=\sigma^2-\sqrt{{(p_1^0)}^2},\;\;\; m_{\pm}=m_{2}\pm m_{3}.
\end{eqnarray*}
Here $M_{{\Xi_{cc}^{++}} }$, $m_{1}$, $m_{2}$ and $m_3$ denote the
masses of the doubly charmed baryon, up and down quarks
respectively. In the following, for obtaining numerical results,
we use the Monte Carlo method to carry out this integral. For
baryon $\Xi_{bb}^-$, the expression is the same but only the mass
of charm quark is replaced by that of bottom quark.

\section{Numerical results}

Now we present our numerical results.

Since only $\Xi_{cc}^{++}$ has been measured, in the later
calculation, we use its measured mass as input, and for
$\Xi_{bb}^-$ we will only illustrate the dependence of its decay
rate on the parameters. The input parameters include: $G_F =
1.166\times10^{-5}\;\rm{GeV^{-2}}$, $m_c = 1.60$ GeV, $m_u = m_d =
0.3$ GeV, $m_s=0.45$ GeV, $m_b=4.87$ GeV. $M_{\Xi_{cc}^+}= 3.519$
GeV. $\alpha_\rho=0.33\;\rm{GeV^2},
\alpha_\lambda=0.25\;\rm{GeV^2}$ \cite{PDG,Wang,SELEX-DOUBLE
CHARM,bbd}. Here, the light quark mass refers to the constitute
mass.

\subsection{The results in the Unparticle scenario}
\begin{table}[htb]
\caption{The decay widths of $\Xi_{cc}(3520)^+ \to uud$ or
$\Xi_{cc}^{++} \to uuu$ and $\Xi_{bb}^-\to ddd(ssd)$ corresponding
to $d_{\mathcal{U}}=3/2$ (in units of GeV). In the table, the
second, third and fourth columns respectively correspond to the
contributions from exchanging scalar unparticle, vector unparticle
and both. \label{typical}}
\begin{center}
\begin{tabular}{c||ccc}
  \hline
  &scalar& vector& scalar+vector\\\hline\hline
  $\Gamma[\Xi_{cc}(3520)^+\to uud]$& $4.57\times 10^{-18}$ &$1.11\times 10^{-15}$
  &$1.24\times 10^{-15}$ \\\hline
  $\Gamma[\Xi_{bb}^-\to ddd]$& $5.85\times 10^{-20}$ &$1.65\times 10^{-17}$
  &$1.83\times 10^{-17}$ \\\hline
  $\Gamma[\Xi_{bb}^-\to ssd]$& $5.21\times 10^{-20}$ &$1.23\times 10^{-17}$
  &$1.44\times 10^{-17}$ \\\hline
  \end{tabular}\label{t1}
\end{center}
\end{table}

For the unknown parameters $\Lambda_\mathcal{U}$ in the unparticle
scenario, according to the general discussion, the energy scale
may be at order of TeV, thus one can fix $\Lambda_\mathcal{U}=1$
TeV. We choose $d_\mathcal{U}=3/2$ in our calculation.

In this work, we also calculate the inclusive decay width of
doubly bottomed baryon $\Xi_{bb}^{-}$. The mass of $\Xi_{bb}^{-}$
is set as $10.09$ GeV according to the estimate of Ref.
\cite{bbd}, although there are no data available yet.

\begin{figure}[htb]
\caption{(a) and (b) respectively show the dependences of the decay
widths of $\Xi_{cc}(3520)^+\to uud$ and $\Xi_{bb}^-\to ddd$
respectively coming from the contributions of scalar unparticle,
vector unparticle and both on $d_{\mathcal{U}}$. \label{ccd-2}}
\begin{center}
\begin{tabular}{c}.
\scalebox{0.75}{\includegraphics{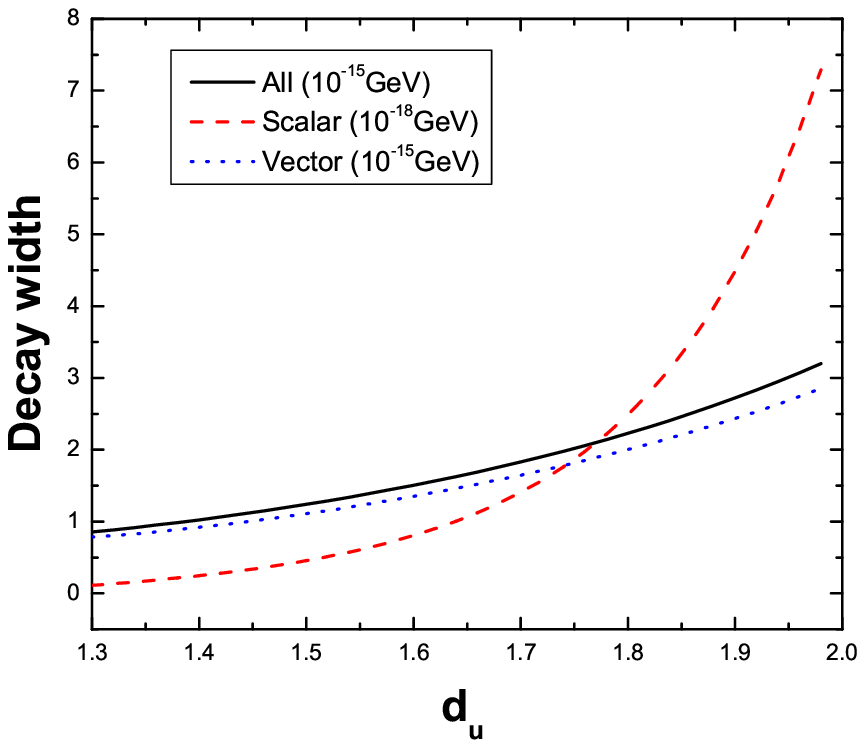}}\\(a)\\
\\
\scalebox{0.75}{\includegraphics{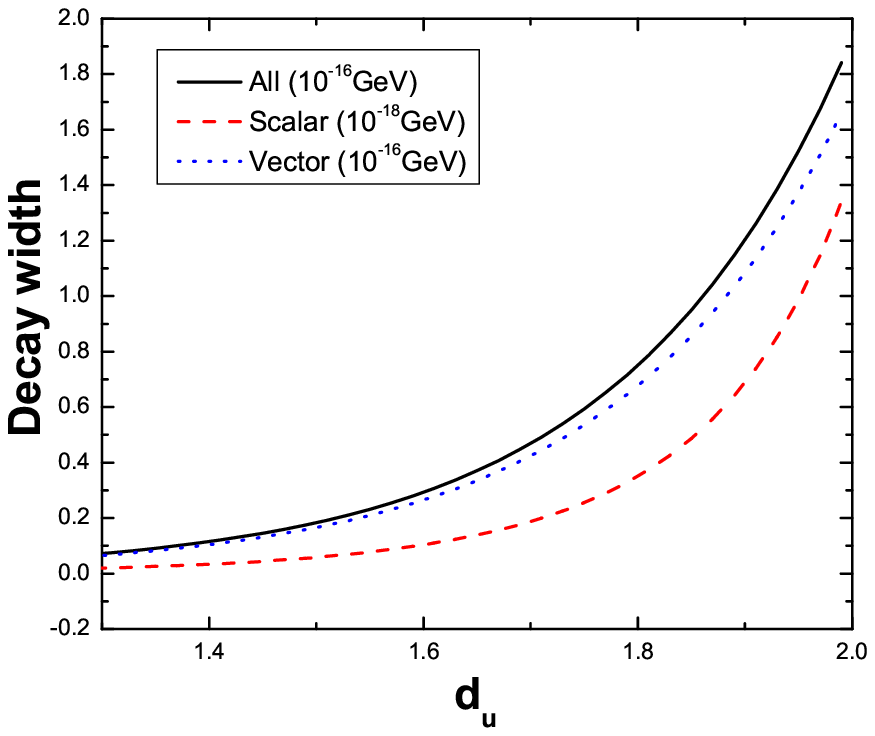}}\\(b)
\end{tabular}
\end{center}
\end{figure}

The numerical results are provided in Table \ref{t1}. Fig.
\ref{ccd-2} illustrates the dependence of the decay widths of
$\Xi_{cc}(3520)^+\to uud$ and $\Xi_{bb}^-\to ddd$ on
$d_{\mathcal{U}}$, the three lines (solid, dashed and dotted)
correspond to the contributions of scalar unparticle, vector
unparticle and both on $d_{\mathcal{U}}$. It is noted here "both"
means that at present we cannot determine whether the unparticle is
a scalar or vector and it is also possible that both scalar and
vector exist simultaneously. Thus we assume both of scalar and
vector contribute and they interfere constructively. Definitely, it
is worth of further investigation.

\subsection{ The results for $Z'$ exchange}

The earlier studies indicate that the mass of $M_{Z'}$ should be
larger than 500 GeV \cite{h1} and $V^{u*}_{Rtc}V^{u*}_{Rtu}$ is
bound  no more than $2.0\times 10^{-4}$ \cite{h2}, in our
calculation, we take their extreme values as $M_{Z'}=500$ GeV and
$V^{u*}_{Rtc}V^{u*}_{Rtu}=2.0\times 10^{-4}$, thus we would obtain
the upper limit of the decay width. It is estimated with all the
input parameters as
\begin{eqnarray}\label{Zresult}
\Gamma[\Xi_{cc}(3520)^+\to uud]=7.66\times 10^{-21}\;\rm{GeV}.
\end{eqnarray}

This is a too small numerical value compared with the width of
$\Xi_{cc}^{+}$, therefore, it is hopeless to observe a non-zero
branching ratio of $\Xi_{cc}^+$ into charmless final states if
only $Z'$ is applied.

\subsection{Estimate the contribution from Standard Model}

As indicated above, in the framework of the SM,  $\Xi_{cc}^+$ can
decay into two-body final states via Fig. \ref{SM} (c). It would
be interesting to compare the SM contribution with that from the
two models. Thus, we would roughly estimate the ratio of the
contribution of Fig. \ref{SM} (c) to that of Fig. \ref{decay} (b)
for the $Z'$ model. It is easier to compare them because the
structures of two diagrams and the relevant effective vertices are
similar.

By the order of magnitude estimation and with the SU(3) symmetry,
the ratio of the amplitude of Fig. \ref{SM} (c) ($T_{SM}$) to Fig.
\ref{decay} (b) ($T_{un}$ or $T_{Z'}$) is
\begin{eqnarray}
\frac{T_{SM}}{T_{Z'}}\approx\frac{8G_F}{\frac{4\pi
\alpha_s}{q^2}\sqrt{2}(V^{u*}_{Rti}V^{u}_{Rtj})^2},
\end{eqnarray}
where $q$ is the momentum of unparticle or $Z'$ (in the case of
the SM, $q^2\ll M_W^2$) and can be neglected in the propagator.
Because the contribution from smaller $q^2$ is dominant, in the
estimation, we set $q^2=0.5 \rm{GeV}^2$. Since the whole case
under consideration, may fall in the non-perturbative QCD region,
as a rough estimate, we take $\alpha_s=1$,
$V^{u*}_{Rti}V^{u}_{Rtj}=2\times 10^{-4}$ and $G_F=1.166\times
10^{-5} \rm{GeV}^{-2}$, we can get $\frac{T_{SM}}{T_{Z'}}\approx
66$ (i.e.$\frac{\Gamma_{SM}}{\Gamma_{Z'}}\approx 4300$ ). Then we
can obtain the ratio of decay widths
$\frac{\Gamma_{Unparticle}}{\Gamma_{SM}}\approx 40$. This ratio
indicates that for $\Xi_{cc}^+$ the contribution of the SM is
smaller than that of the unparticle scenario, but larger than that
from the $Z'$ model.

However, for $\Xi_{cc}^{++}$, Fig.\ref{SM}(c) does not contribute
at all, so that the decay  of $\Xi_{cc}^{++}$ into charmless final
states (or $\Xi_{bb}^-$ into bottomless final states) is more
appropriate for exploring new physics than $\Xi_{cc}^{+}$.

It is worth noticing that the estimate of the contribution of the
SM to the decay rate is very rough, thus what we can assure to
ourselves is its order of magnitude. Indeed the magnitude
contributed by the SM is very small and cannot produce sizable
observational effects at all, even though it has a comparable
order with that from the two sample models, the unparticle and
$Z'$. In the future, if such mode were observed at LHCb, we can
definitely conclude that it is not caused by the SM, but new
physics.

\section{Discussion and conclusion}

In this work, we propose to explore for new physics beyond the SM
at LHCb by measuring direct decays of $\Xi_{cc}^{+}\;
\Xi_{cc}^{++},\; (\Xi_{bb}^-,\;\Xi_{bb}^0) $ into charmless
(bottomless) final states. Such decays can occur via the diagrams
shown in Fig. 1 in the framework of the SM, but is much suppressed
to be experimentally observed, therefore if a sizable rate is
measured, it would be a clear signal for new physics beyond the
SM. We use two models as examples, namely the unparticle and $Z'$
models to calculate the decay rates, because both of them allow a
transition of $cc\;(bb)\to qq$ where $q$ may be light quarks to
occur at tree level. Thus one expects that these new models might
result in non-zero observation.

Indeed, our work is motivated by three factors, first the great
machine LHC will run next year and a remarkable amount of data
will be available, then secondly, the double-charmed baryon
$\Xi_{cc}$ which was observed by the SELEX collaboration provides
us a possibility to probe new physics, and the last reason is that
some models have been proposed and they may induce a
flavor-changing neutral current, concretely the unparticle and
$Z'$ models are employed in this work. Definitely none of the two
models are confirmed by either theory or experiment yet, and they
still need further theoretical investigations, but their framework
is clear, so that we may use them as examples to demonstrate how
new physics may cause such decay modes and indicate that a sizable
observational rate is a clear signature for new physics beyond the
SM. Moreover, the double charmed baryon has only been observed by
the SELEX collaboration, but not at B-factories.  It seems
peculiar at first glimpse, but careful studies indicate that it is
quite reasonable due to the fragmentation process of heavy quarks.
The authors of Ref.\cite{Chang} indicate that the meson $B_c$
cannot be seen at any $e^+e^-$ colliders because its production
rate at such machines is too small, but by contraries, its
production rate is greatly enhanced at hadron colliders. It was
first observed at TEVATRON and its production rate at LHC would be
much larger by several orders \cite{Chang}. In analog, one can
expect that such double-charmed baryons $\Xi_{cc}$ or
double-bottomed $\Xi_{bb}$ can only be produced at LHC, but not at
B-factories.

The inclusive decays of doubly charmed baryons $\Xi_{cc}(3520)^+$,
$\Xi_{cc}^{++}$ and $\Xi_{bb}^0$, $\Xi_{bb}^-$ are explored in
unparticle and $Z'$ scenarios. Our result indicates that the upper
limit of the inclusive decay width of $\Xi_{cc}^{++}\to uuu$ is
about $10^{-15}$ GeV with $d_{\mathcal{U}}=3/2$. For inclusive
decay $\Xi_{bb}^-\to ddd(ssd)$, the upper limit is at order of
$10^{-17}$ GeV. It is learnt that in the unparticle scenario, the
contribution from exchanging a vector unparticle is much larger
than that from exchanging a scalar unparticle, as shown in Table
\ref{typical}.

The parameters which we employ in the numerical computations are
obtained by fitting other experimental measurements, for example
if the recently observed $D^0-\bar D^0$ can be interpreted by the
unparticle model, an upper bound on the parameters in the model
would be constrained. Indeed, all the present experimental data
can only provide upper bounds on the model parameters no matter
what new physics model under consideration is.

So far it is hard to make an accurate estimate on the production
rates of the heavy baryons which contain two heavy quarks at LHC
yet, but one has reason to believe that the production rate would
be roughly of the same order of the production rate of $B_c$ which
was evaluated by some authors \cite{Chang}, or even smaller by a
factor of less than 10. The production rates indeed will be
theoretically evaluated before or even after LHC begins running.

In Ref. \cite{production rate}, the authors estimate the number of
$\Xi_{cc}$ produced at LHCb as about $10^9$. Since the available
energy is much higher than the masses of $\Xi_{cc}$ and
$\Xi_{bb}$, one has strong reason to believe that their production
rates are comparable. Unfortunately our numerical results indicate
that the unparticle and $Z'$ scenarios cannot result in sizable
rates for $\Xi_{cc}^{++}\to uuu\to\; two\; hadrons$ and
$\Xi_{bb}^-\to ddd(ssd)\to\; two\; hadrons$ which can be measured
at LHCb and neither the SM. Even though the two sample models and
SM cannot cause sufficiently large rates, the channels still may
stand for a possible place to search for new physics. If a sizable
rate is observed at LHCb, it would be a signal of new physics and
the new physics is also not the unparticle and/or $Z'$, but
something else.

\section*{Acknowledgments}

We greatly benefit from constructive discussions with Prof.
Xiao-Gang He. This project was supported by the National Natural
Science Foundation of China under Grants 10475042, 10721063,
10625521, 10705001, 10745002 and 10705015, Key Grant Project of
Chinese Ministry of Education (No. 305001), Ph.D. Program
Foundation of Ministry of Education of China and the China
Postdoctoral Science foundation (No. 20060400376).

%\end{CJK}
\end{document}